\documentclass[print]{revtex4}
\usepackage{amsmath,amssymb}
\usepackage{graphicx}

\begin{document}

\title{\large \bf  Scheme for conditional generation of photon-added coherent state and optical entangled $W$ state}

\author{Yan Li,$^{1,2,3}$\footnote{Electronic address: liyan@wipm.ac.cn} Hui Jing,$^{1,2}$\footnote{Electronic address: jinghui@wipm.ac.cn} and Ming-Sheng Zhan$^{1,2}$\footnote{Electronic address: mszhan@wipm.ac.cn}}

\affiliation{$^{1}$State Key Laboratory of Magnetic Resonance and
Atomic and Molecular Physics,\\
 Wuhan Institute of Physics and Mathematics, Chinese Academy of Sciences, Wuhan 430071, P. R. China\\
 $^{2}$Center for Cold Atom Physics, Chinese Academy of Sciences, Wuhan
 430071, P. R. China\\
$^{3}$Graduate School of the Chinese Academy of Sciences, Beijing
100080, P. R. China}

\begin{abstract}
We propose a simple scheme to generate an arbitrary photon-added
coherent state of a travelling optical field by using only a set
of degenerate parametric amplifiers and single-photon detectors.
Particularly, when the single-photon-added coherent state (SPACS)
is observed by following, e.g., the novel technique of Zavatta
\emph{$et~al.$} (Science 306, 660 (2004)), we also obtain the
generalized optical entangled $W$ state. Finally, a qualitative
analysis of possible losses in our scheme is given.\\

PACS numbers: 03.67.Mn, 42.50.Dv, 03.65.Ud
\end{abstract}

\baselineskip=16pt

\maketitle

$$\texttt{I~~INTRODUCTION}$$

\indent The generation and engineering of quantum state play a key
role in current quantum information science. Over the past decade,
various schemes of preparations of different kinds of nonclassical
or entangled quantum states have been probed by using, e.g., the
versatile nonlinear medium or the conditional measurement
technique at the output ports of beam splitters (BS)\cite{1}. For
some elegant examples, Sanders proposed a scheme to produce an
entangled coherent state by using a nonlinear Kerr medium in one
arm of an interferometer \cite{2}; Dakna \emph{$et~al.$} proposed
a method relying on an alternate application of coherent
displacement and photon adding or subtracting via conditional
measurements on BS for the generation of several different types
of nonclassical states \cite{3}; Lvovsky \emph{$et~al.$} designed
a scheme to prepare the highly nonclassical displaced Fock states
of harmonic oscillators by acting upon Fock states with
displacement operators \cite{4}; Sanaka \cite{5} and Pegg
\emph{$et~al.$} \cite{6} demonstrated the production of
nonclassical Fock state of light or its finite superpositions by
using the effective nonlinearity of BS or truncating a classical
coherent state; and the three- or four-photon states also were
achieved by using correlated photon pairs produced by parametric
down conversion and single-photon detector (SPD) with the
capability to discriminate photon-number state \cite{7}.

\indent Recently, Agarwal and Tara formulated an interesting new
nonclassical quantum state, i.e., the photon-added coherent state
(PACS), which exhibits the intermediate properties between the
classical coherent state and the quantum Fock state, and also gave
a possible scheme to produce this state via cavity QED \cite{8}.
In their remarkable experiment, comprised of type-I beta-barium
borate(BBO)crystal, SPD and balanced homodyne detector, Zavatta
\emph{$et~al.$} obtained the novel single-photon-added coherent
state (SPACS) and then firstly visualized the intriguing
transition process from classical to purely quantum regimes
\cite{9}. In this paper, following these important pioneering
works, we propose a simple but new scheme to produce an arbitrary
PACS by combining an array of parametric amplifiers and
corresponding SPDs. Particularly, the generalized optical
$N$-qubit entangled $W$ state, as an important resource for
current quantum information science \cite{10,11,12,13,14}, also
can be generated probabilistically in idler channels when we get
the novel SPACS in output signal channel.

$$\texttt{II~~THE~~THEORETICAL~~MODEL}$$

\indent The PACS $|\alpha,m\rangle$, firstly introduced by Agarwal
and Tara \cite{8}, is defined as
\begin{equation}\label{eqn:1}
|\alpha,m\rangle=\frac{\hat{a}^{\dag
m}|\alpha\rangle}{[m!L_{m}(-|\alpha|^{2})]^{1/2}},
\end{equation}
where $\hat{a}(\hat{a}^{\dag})$ is the photon annihilation
(creation) operator, $m$ is an integer, $L_{m}(x)$ is the Laguerre
polynomial of order $m$ defined by
$L_{m}(x)=\sum_{n=0}^{m}\frac{(-1)^{n}x^{n}m!}{(n!)^{2}(m-n)!}$.
Obviously, when $\alpha\rightarrow0$ or $m\rightarrow0$,
$|\alpha,m\rangle$ reduces to the Fock or coherent state
respectively. The novel properties of PACS, as an intermediate
state between the quantum and classical limits, was studied in
detail by Agarwal \emph{$et~al.$} \cite{8}. Note that it is quite
different from another intermediate state, i.e., the displaced
Fock state:
$|DFS\rangle=D(\alpha)|n\rangle=exp(\hat{a}^{\dag}\alpha-\hat{a}\alpha^{*})|n\rangle$,
which is generated by acting upon Fock states with displacement
operators \cite{4}. The PACS is, however, obtained via successive
one-photon excitations on a classical coherent light.

\indent Now we consider the parametric down-conversion process
(type-I BBO crystal) as an optical parametric amplifier, in which
one photon incident on the dielectric with $\chi^{2}$ nonlinearity
breaks up into two new photons of lower frequencies. In the steady
state, we always have $\omega_{0}=\omega_{1}+\omega_{2}$, with
$\omega_{0}$ the pump frequency, and $\omega_{1}$, $\omega_{2}$
the signal and idler frequencies. Under the phase matching
condition, the wave vectors of the pump, signal and idler photons
are related by $\vec{k}_{0}=\vec{k}_{1}+\vec{k}_{2}$ (momentum
conservation). The signal and idler photons appear simultaneously
within the resolving time of the detectors and the associated
electronics \cite{15}. The Hamiltonian of this process can be
written as:
\begin{equation}\label{eqn:3}
\hat{H}=\sum_{i=0}^{2}\hbar\omega_{i}(\hat{n}_{i}+\frac{1}{2})+\hbar
g[\hat{a}_{1}^{\dag}\hat{a}_{2}^{\dag}\hat{a}_{0}+\hat{a}_{1}\hat{a}_{2}\hat{a}_{0}^{\dag}],
\end{equation}
where the real mode coupling constant $g$ contains the nonlinear
susceptibility $\chi^{2}$. Besides,
$[\hat{n}_{1}+\hat{n}_{2}+2\hat{n}_{0}, \hat{H}]= 0$, which
indicates the conversion of one pump photon into one signal and
one idler photon.

\indent For simplicity, we suppose that signal and idler waves
have the same polarizations and that their directions are well
defined by apertures. Also the incident pump field is intense and
the pump mode $\hat{a}_{0}$ can be treated classically as a field
with complex amplitude $\hat{a}_{0}\rightarrow iV$(see Fig.1).
Then the Hamiltonian in the interaction picture is:
$\hat{H}_{I}=i\hbar
gV[\hat{a}_{1}^{\dag}\hat{a}_{2}^{\dag}-\hat{a}_{1}\hat{a}_{2}]$
and, for an initial state:
$|\psi(0)\rangle=|\alpha\rangle_{s}|0\rangle_{i}$, after an
interaction time $t$, the output state evolves into
\begin{equation}\label{eqn:1}
|\psi(t)\rangle=exp[-i\hat{H}_{I}t/\hbar]|\psi(0)\rangle
=exp[\lambda(\hat{a}_{1}^{\dag}\hat{a}_{2}^{\dag}-\hat{a}_{1}\hat{a}_{2})]|\psi(0)\rangle,
\end{equation}
where $\lambda=Vgt$, as an effective interaction time. For short
times $t$ compared with average time interval between successive
down-conversions, by expansion of the exponential we have
(assuming $\lambda\ll 1$)
\begin{equation}\label{eqn:1}
|\psi(t)\rangle
\approx|\alpha\rangle_{s}|0\rangle_{i}+\lambda|\alpha,1\rangle_{s}|1\rangle_{i}+
\frac{\lambda^{2}}{2}(\hat{a}_{1}^{\dag}\hat{a}_{2}^{\dag}\hat{a}_{1}^{\dag}\hat{a}_{2}^{\dag}-
\hat{a}_{1}\hat{a}_{2}\hat{a}_{1}^{\dag}\hat{a}_{2}^{\dag})|\alpha\rangle_{s}|0\rangle_{i}.
\end{equation}
Since $\lambda\ll 1$, we can select the first two terms as the
output state, i.e., $
|\psi(t)\rangle\approx|\alpha\rangle_{s}|0\rangle_{i}+\lambda|\alpha,1\rangle_{s}|1\rangle_{i}
$.

So far, the output signal and idler lights are entangled with each
other in the frequency domain \cite{15}. From the above we can see
that the signal channel mostly contains the original coherent
state, except for the few cases when single photon is detected in
output idler channel. Thereby these rare events can stimulate
emissions of one photon into the coherent state, which then
generates the novel SPACS in the signal channel with a success
probability being proportional to $|\lambda|^{2}(1+|\alpha|^{2})$.
\begin{figure}[ht]
\includegraphics[width=0.6\columnwidth]{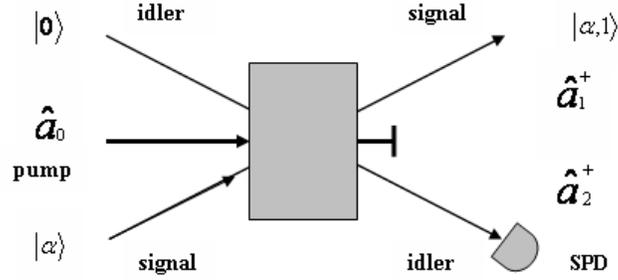}
\caption{The process of single-photon excitation of the nonlinear
crystal in the signal coherent channel. The pump light is the
classical coherent light of high intensity $|V|^{2}$, and the
idler input channel is vacuum. SPD is the single photon detector
placed in the idler channel. When SPD detects one photon, the
coherent state will be excited by the entangled photon-pairs in
the signal channel.} \label{}
\end{figure}

Note that, recently, Zavatta \emph{$et~al.$} first experimentally
created the SPACS and then visualized the novel quantum-classical
transition process via the technique of single-photon detection
and balanced homodyne detection \cite{9,16}. Here we select the
one-photon excitation term to avoid higher-order ones which cannot
be discriminated by the SPD in the elegant experiment of Zavatta
\emph{$et~al.$}\cite{16}.

$$\texttt{III~~PREPARATION~~OF~~THE~~DESIRED~~STATES}$$

\indent Turning to Fig.2, we combine two identical optical
parametric amplifiers by assuming the same low-gain regime, i.e.,
$g_{1}=g_{2}=g$ and the same strong classical pumps, i.e.,
$iV_{1}=iV_{2}=iV$. Therefore, for an initial input state: $
|\psi(0)\rangle=|\alpha\rangle_{s10}|0\rangle_{i10}|0\rangle_{i20}
$, after some evolution time $t$, the output state becomes
\begin{eqnarray}
&&|\psi(t)\rangle\nonumber
=exp[-i\hat{H}_{I2}t_{2}/\hbar]exp[-i\hat{H}_{I1}t_{1}/\hbar]|\psi(0)\rangle\\
&& \approx
|\alpha\rangle_{s2}|0\rangle_{i1}|0\rangle_{i2}+\lambda_{1}|\alpha,1\rangle_{s2}|1\rangle_{i1}|0\rangle_{i2}
+\lambda_{2}|\alpha,1\rangle_{s2}|0\rangle_{i1}|1\rangle_{i2}+
\lambda_{1}\lambda_{2}|\alpha,2\rangle_{s2}|1\rangle_{i1}|1\rangle_{i2}.
\end{eqnarray}\label{eqn:1}
Here, $\lambda_{j}=Vgt_{j} (j=1, 2)$, denoting different effective
time for different crystals. For this, there are three cases based
on the results of SPDs in idler channels:

\noindent

\begin{figure}[ht]
\includegraphics[width=0.6\columnwidth]{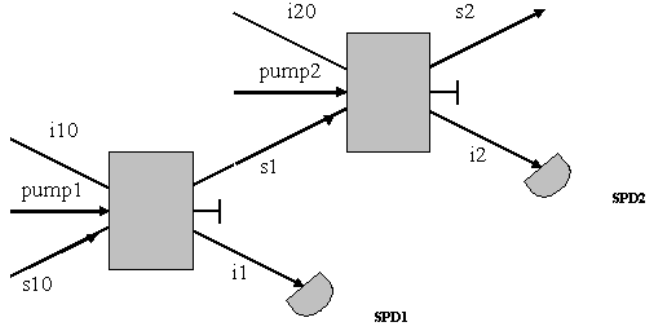}
\caption{The scheme to generate $|\alpha,1\rangle$ and
$|\alpha,2\rangle$ by combining two same nonlinear crystals with
two same pump lights. Given the coherent light in signal channel
and vacuum in two idler channels input, the desired states will be
generated with different probability on conditional detections of
SPD1 and SPD2.} \label{}
\end{figure}

Case 1: Neither of two detectors catches one photon. The initial
input coherent state just appears in output signal channel most of
the time.

Case 2: One of the detectors catches one photon and the other does
not. The SPACS can be created in output signal channel in several
cases with a success probability being proportional to:
$(|\lambda_{1}|^{2}+|\lambda_{2}|^{2})(1+|\alpha|^{2})$.

Case 3: Both the detectors detect one photon simultaneously. There
will be a double-photon-added coherent state (DPACS)
$|\alpha,2\rangle$ appearing in output signal channel in the quite
rare cases with a success probability being proportional to:
$|\lambda_{1}\lambda_{2}|^{2}2!L_{2}(-|\alpha|^{2})$.

In short, the states $|\alpha,1\rangle$, $|\alpha,2\rangle$ can
be probabilistically prepared via the conditional SPDs technique.\\

\indent Similarly we can combine three identical optical
parametric amplifiers as in Fig.3, and now, for an initial input
state of the three-body system:
$|\psi(0)\rangle=|\alpha\rangle_{s10}|0\rangle_{i10}|0\rangle_{i20}|0\rangle_{i30}$,
after an interaction time $t$, the output state of the system is
written as
\begin{eqnarray}
&&|\psi(t)\rangle\nonumber
=exp[-i\hat{H}_{I3}t_{3}/\hbar]exp[-i\hat{H}_{I2}t_{2}/\hbar]exp[-i\hat{H}_{I1}t_{1}/\hbar]|\psi(0)\rangle\\\nonumber
&& \approx
|\alpha\rangle_{s3}|0\rangle_{i1}|0\rangle_{i2}|0\rangle_{i3}+\lambda_{1}|\alpha,1\rangle_{s3}|1\rangle_{i1}|0\rangle_{i2}|0\rangle_{i3}+\lambda_{2}|\alpha,1\rangle_{s3}|0\rangle_{i1}|1\rangle_{i2}|0\rangle_{i3}+\lambda_{3}|\alpha,1\rangle_{s3}|0\rangle_{i1}|0\rangle_{i2}|1\rangle_{i3}\\\nonumber
&&+\lambda_{1}\lambda_{2}|\alpha,2\rangle_{s3}|1\rangle_{i1}|1\rangle_{i2}|0\rangle_{i3}+\lambda_{2}\lambda_{3}|\alpha,2\rangle_{s3}|0\rangle_{i1}|1\rangle_{i2}|1\rangle_{i3}+\lambda_{1}\lambda_{3}|\alpha,2\rangle_{s3}|1\rangle_{i1}|0\rangle_{i2}|1\rangle_{i3}\\
&&+\lambda_{1}\lambda_{2}\lambda_{3}|\alpha,3\rangle_{s3}|1\rangle_{i1}|1\rangle_{i2}|1\rangle_{i3},
\end{eqnarray}\label{eqn:1}
in which, $\lambda_{j}=Vgt_{j} (j=1, 2, 3)$, denoting different
evolution times for different nonlinear crystals.
\begin{figure}[ht]
\includegraphics[width=0.8\columnwidth]{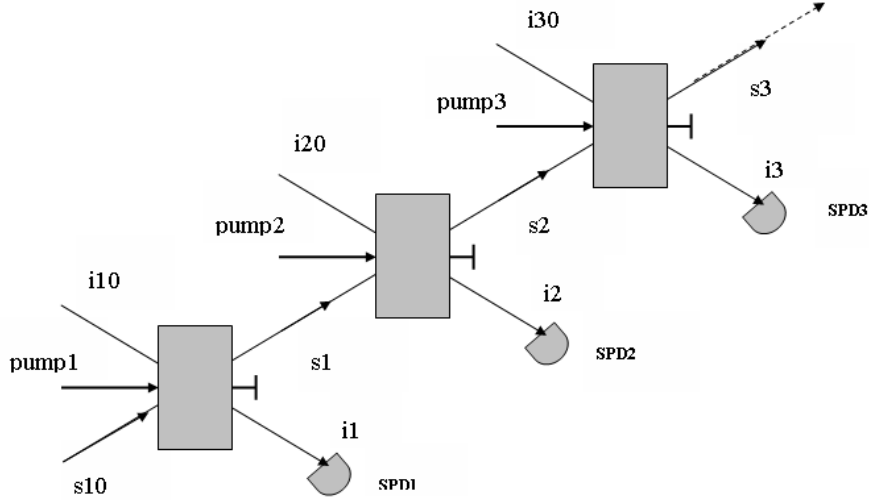}
\caption{The scheme to generate $|\alpha, m\rangle$ by combining a
series of same nonlinear crystals with all the pump lights equal.
Given the coherent light in signal channel and vacuum in all of
the idler channels input, arbitrary PACS will be generated with
different probability in principle on conditional detections by
the corresponding series of SPDs. In addition, the multi-qubit $W$
state will be produced probabilistically in output idler channels
on conditional generation of SPACS in output signal channel.}
\label{}
\end{figure}

From the above output state of the three-body system, we can
probabilistically get the PACS in output signal channel on
conditional detections of the single-photon in three idler
channels, i.e.,

Case 1: Neither of three SPDs detects one photon. Again we get a
coherent output signal $|\alpha\rangle_{s3}$.

Case 2: When one SPD detects one photon and the other two cannot.
The SPACS can be generated in output signal channel with a success
probability being proportional to:
$(|\lambda_{1}|^{2}+|\lambda_{2}|^{2}+|\lambda_{3}|^{2})(1+|\alpha|^{2})$.

Case 3: When two SPDs detect one photon simultaneously and the
third cannot. Now we get the DPACS $|\alpha,2\rangle_{s3}$ in
output signal channel but with a success probability being
proportional to:
$(|\lambda_{1}\lambda_{2}|^{2}+|\lambda_{2}\lambda_{3}|^{2}+|\lambda_{1}\lambda_{3}|^{2})2!L_{2}(-|\alpha|^{2})$.

Case 4: When the three SPDs all detect one photon instantaneously.
There will be a triple-photon-added coherent state (TPACS)
$|\alpha,3\rangle_{s3}$ appearing in output signal channel with a
success probability being proportional to:
$|\lambda_{1}\lambda_{2}\lambda_{3}|^{2}3!L_{3}(-|\alpha|^{2})$.

We should note that the most interesting thing comes when we get
the novel state SPACS in output signal channel, i.e., the final
output state will be the following entangled state:
\begin{equation}\label{eqn:1}
\lambda_{1}|1\rangle_{i1}|0\rangle_{i2}|0\rangle_{i3}+
\lambda_{2}|0\rangle_{i1}|1\rangle_{i2}|0\rangle_{i3}+\lambda_{3}|0\rangle_{i1}|0\rangle_{i2}|1\rangle_{i3},
\end{equation}
for the equal effective time, i.e.,
$\lambda_{1}=\lambda_{2}=\lambda_{3}=\lambda$, this state can be
written as
\begin{equation}\label{eqn:1}
|\psi(t)\rangle_{W}^{3}=\lambda(|1\rangle|0\rangle|0\rangle+|0\rangle|1\rangle|0\rangle+|0\rangle|0\rangle|1\rangle),
\end{equation}
with the success probability $
P_{suc}^{W}\propto|\lambda|^{2}(1+|\alpha|^{2}). $ Obviously, if
we choose a proper effective interaction time, i.e.,
$\lambda=\frac{1}{\sqrt{3}}$, we can obtain an interesting
entangled state which takes the same form as the familiar
three-qubit entangled $W$ state: $ |W\rangle=\frac{1}{\sqrt{3}}(|1
0 0\rangle+|0 1 0\rangle+|0 0 1\rangle). $

\indent The proposed generation scheme can be used to the case of
$|\alpha, m\rangle$ and $N$-qubit entanglement in a
straightforward way. The scheme is as Figure 3 shows with a series
of such combinations. Assuming a system composed of $N$ same pump
lights and $N$ same amplifiers with the input signal channel
coherent light and the rest idler channels vacuum light, the input
initial state reads:
\begin{equation}\label{eqn:1}
|\psi(0)\rangle=|\alpha\rangle_{s10}|0\rangle_{i10}|0\rangle_{i20}\ldots|0\rangle_{iN0}.
\end{equation}
After a series of Hamiltonian interaction, the output state of the
$N$-body system reads:
\begin{equation}\label{eqn:1}
|\psi(t)\rangle
=\prod_{j=1,N}exp[-i\hat{H}_{Ij}t_{j}/\hbar]|\psi(0)\rangle,
\end{equation}
where
\begin{equation}\label{eqn:1}
\hat{H}_{Ij}=i\hbar g
V[\hat{a}_{1j}^{\dag}\hat{a}_{2j}^{\dag}-\hat{a}_{1j}\hat{a}_{2j}].
\end{equation}

After an evolution time $t$, the output entangled state of the
N-body system can be got from the Eq.(10). Based on different
conditional detections of SPDs in the idler channels, arbitrary
$|\alpha,m\rangle$ can be generated with the success probability
being proportional to $N|\lambda^{m}|^{2}m!L_{m}(-|\alpha|^{2})$
with $m<N$, $t_{j}=\frac{t}{N}$, in which $t_{j}$ is the real
interaction time with each medium.

At the same time, the $N$-qubit $W$ state in the idler channels
will be produced on conditional generation of the SPACS in the
output signal channel. When we choose
$\lambda=\frac{1}{\sqrt{N}}$, we can get:
\begin{equation}\label{eqn:1}
|\psi(t)\rangle_{W}^{N}=\frac{1}{\sqrt{N}}(|1\underbrace{00\cdots0}_{N-1}\rangle+|01\underbrace{00\cdots0}_{N-2}\rangle+\cdots+|\underbrace{00\cdots0}_{N-1}1\rangle).
\end{equation}
For simplicity, we rewrite it as
\begin{equation}\label{eqn:1}
|\psi(t)\rangle_{W}^{N}=\frac{1}{\sqrt{N}}|N-1,1\rangle,
\end{equation}
where $|N-1,1\rangle$ denotes all the totally symmetric states
involving $N-1$ zeros and $1$ one. It follows that, the success
probability of $N$-qubit $W$ state is: $
P_{suc}^{W}\propto|\lambda|^{2}(1+|\alpha|^{2}). $

$$\texttt{IV~~DISCUSSIONS}$$

We propose a simple scheme to generate an arbitrary PACS, at the
same time, the optical entangled $W$ state can be created on the
detection of the SPACS in output signal channel. In addition, if
the TPACS $|\alpha,2\rangle$ can be efficiently detected, one also
can get the entangled $W$ state with a quadratic damping
probability. The three-qubit $W$ state, as one of the two
inequivalent classes of genuine tripartite entangled state (i.e.,
the GHZ \cite{17,18} and the $W$ states \cite{19,20}), shows
perfect correlations and violates a three-partite Mermin
inequality, though its violation is weaker than that for the GHZ
state \cite{21}. The most interesting properties of $W$ state is
that if one particle is measured in basis of
{$|0\rangle$,$|1\rangle$}, then the state of remained two
particles is either in a maximally entangled state or in a product
state, which has important applications in current quantum
information science.

Differing from the scheme of generating three-photon
polarization-entangled $W$ state proposed by Eibl \emph{$et~al.$}
based on polarization measurements and type-II spontaneous
parametric down-conversion \cite{21}, our scheme needs the
technique of SPDs and the SPACS reconstructions. Therefore, if the
efficiency of the SPDs and the reconstruction accuracy of SPACS is
high enough, the optical entangled $W$ state can be generated
efficiently. In the elegant experiment of Zavatta \emph{$et~al.$}
\cite{16}, a high-frequency time-resolved balanced homodyne
detection was used to reconstruct the Wigner function of SPACS
with an overall efficiency of $60\%$. Hence we can use this
homodyne technique to enhance the efficiency of $W$ state
generation. Besides, since the $W$ state can be created whenever
one SPD catches one photon, the SPDs with high quantum efficiency
should be used to get a high generation rate.

It should be emphasized that the probability of $W$ state
generation is independent of crystal numbers $N$, and is
determined only by the input seed coherent light and the effective
interaction time. And higher generation rate can be got by, at
least theoretically, enhancing the intensity of pump light and
input coherent light and enlarging the couplings with the medium.
However, there are always some realistic problems related to the
imperfection of the elements itself which affects the generation
efficiency and the fidelity of the desired output state, and many
authors analyzed the losses of such optical elements as well as
the suggestions of improving the efficiency and fidelity
\cite{22,23}.

One of the difficulties consists in the requirement of high
quantum efficiency of photon-counting detectors. The practical
limitations of the detectors including the nonzero dark count
rates causing false alarm even without a photon in signal mode,
the dead time, during which detectors cannot respond to the
incoming photons, may result in the non-unit quantum efficiency.
In practice, we should choose the detector which bears lower dark
count rate and shorter resolution time on the premise of same
efficiency. Besides, taking into account of the detailed analysis
about the dependence of fidelity on the coherent light intensity
\cite{22}, we can get higher fidelity by lowering the intensity of
the input coherent light. The other difficulty comes from the
non-ideal BBO crystals. Because the photon may have finite
bandwidth due to the finite crystal size and the spatial location
of the idler-signal photons in the radiation cone of the crystal
output, narrow spatial and frequency filters should be placed in
the idler mode before the detector. Since the mode-locked laser,
used as the pump of BBO crystal in Zavatta \emph{$et~al.$}
experiment \cite{9,16}, can provide a quadratic conversion rate
higher than $50\%$ in market, we may select it to enhance the
efficiency.

\indent Since there are other factors causing losses in practice,
i.e., the damping associated with the cascaded system due to the
environment \cite{24}, etc., the problem is more complex and
therefore the successful probability of the desired states is
limited. However, as it is known to all that the generation of
multi-qubit entanglement is difficult in practice for a long time
\cite{25}, our scheme will be experimentally challenging and
promising with the development of SPD and homodyne detection
techniques.
\bigskip

\noindent This work was supported by NSFC (Grant Nos. 10304020 and
10474117), National Basic Research Program of China (Grant No.
2001CB309309), and Wuhan Chenguang Youth Project.







\begin{thebibliography}{99}
\bibitem{1} K. Sanaka, K. J. Resch, and A. Zeilinger, Phys. Rev. Lett. \textbf{96}, 083601 (2006);
            A. P. Lund, H. Jeong, T. C. Ralph, and M. S. Kim, Phys. Rev. A \textbf{70}, 020101(R) (2004);
            H. Jeong, Phys. Rev. A \textbf{72}, 034305 (2005);
            B. M. Escher, A. T. Avelar, and B. Baseia, Phys. Rev. A \textbf{72}, 045803 (2005);
            C. C. Gerry and R. A. Campos, Phys. Rev. A \textbf{64}, 063814 (2001);
            Y. Li, H. Jing and M.-S. Zhan, J. Phys. B: At. Mol. Opt. Phys. \textbf{39}, 2107 (2006).
\bibitem{2} B. C. Sanders, Phys. Rev. A  \textbf{45}, 6811 (1992).
\bibitem{3} M. Dakna, J. Clausen, L. Kn\"{o}ll, and D.-G. Welsch, Phys. Rev. A \textbf{59}, 1658 (1999).
\bibitem{4} A. I. Lvovsky and S. A. Babichev, Phys. Rev. A \textbf{66}, 011801 (2002).
\bibitem{5} K. Sanaka, Phys. Rev. A \textbf{71}, 021801(R) (2005).
\bibitem{6} D. T. Pegg, L. S. Phillips, and S. M. Barnett, Phys. Rev. Lett. \textbf{81}, 1604 (1998).
\bibitem{7} E. Waks, E. Diamanti, and Y. Yamamoto, e-print quant-ph / 0308055.
\bibitem{8} G. S. Agarwal and K. Tara, Phys. Rev. A  \textbf{43}, 492 (1991).
\bibitem{9} A. Zavatta, S. Viciani, and M. Bellini, Science  \textbf{306}, 660 (2004).
\bibitem{10} M. Murao, D. Jonathan, M. B. Plenio, and V. Vedral, Phys. Rev. A \textbf{59}, 156 (1999);
             D. Bru\ss, \emph{et al.,} Phys. Rev. A \textbf{57}, 2368 (1998);
             C. W. Zhang, C. F. Li, Z. Y. Wang, G. C. Guo, Phys. Rev. A \textbf{62}, 042302 (2000).
\bibitem{11} J. Lee, H. Min and S. D. Oh, Phys. Rev. A \textbf{66} 052318 (2002);
             A. Karlsson and M. Bourennane, Phys, Rev. A \textbf{58}, 4394 (1998);
             J. W. Pan, M. Daniell, S. Gasparoni, G. Weihs, A. Zeilinger, Phys. Rev. Lett. \textbf{86} 4435 (2001).
\bibitem{12} J. C. Hao, C. F. Li, G. C. Guo, Phys. Rev. A \textbf{63}, 054301 (2001);
             X. S. Liu, G. L. Long, D. M. Tong, and L. Feng, Phys. Rev. A \textbf{65}, 022304 (2002);
             A. Grudka, A. Wojcik, Phys. Rev. A \textbf{66}, 014301 (2002).
\bibitem{13} S. Bagherinezhad and V. Karimipour, Phys. Rev. A \textbf{67}, 044302 (2003);
             V. Scarani and N. Gisin, Phys. Rev. Lett. \textbf{87}, 117901 (2001).
\bibitem{14} D. Bouwmeester, J.-W. Pan, M. Daniell, H. Weinfurter, and A. Zeilinger, Phys. Rev. Lett. \textbf{82}, 1345 (1999);
             M. Eibl, \emph{et al.}, Phys. Rev. Lett. \textbf{90}, 200403 (2003);
             Z. Zhao, \emph{et al.}, Phys. Rev. Lett. \textbf{91}, 180401 (2003);
             X. L. Feng, Z. M. Zhang, X. D. Li, S. Q. Gong, Z. Z. Xu, Phys. Rev. Lett. \textbf{90}, 217902 (2003);
             J. W. Pan, C. Simon, C. Brukner and A. Zeilinger, Nature \textbf{410}, 1067 (2001).
\bibitem{15} E. Wolf and L. Mandel, Optical Coherence and Quantum Optics, Cambridge University Press (1995).
\bibitem{16} A. Zavatta, S. Viciani, and M. Bellini, Phys. Rev. A \textbf{72}, 023820 (2005)
\bibitem{17} D. M. Greenberger, M. A. Horne, and A. Zeilinger, in
             Bell's Theorem, Quantum Theory, and Conceptions of
             the Universe, edited by M. Kafatos (Kluwer, Dordrecht, 1989), p. 69.
\bibitem{18} J.-W. Pan \emph{et al.,} Nature (London) \textbf{403}, 515 (2000);
             D. Bouwmeester, J. W. Pan, M. Daniell, H. Weinfurter, A. Zeilinger, Phys. Rev. Lett. \textbf{82}, 1345 (1999);
             M. Eibl, \emph{et al.}, Phys. Rev. Lett. \textbf{90}, 200403 (2003).
\bibitem{19} W. D\"{u}r, G. Vidal, and J. I. Cirac, Phys. Rev. A \textbf{62}, 062314 (2000).
\bibitem{20} A. Zeilinger, M. A. Horne, and D. M. Greenberger, in workshop on Squeezed
             States and Uncertainty Relations, edited by D. Han, \emph{et al.},
              NASA conference Publication No. \textbf{3135} (NASA, Washington. DC, 1992), p. 73.
\bibitem{21} M. Eibl, N. Kiesel, M. Bourennane, C. Kurtsiefer, H. Weinfurter, Phys. Rev. Lett. \textbf{92}, 077901 (2004).
\bibitem{22} S. K. \"{O}zdemir, A. Miranowicz, M. Koashi, and N. Imoto, Phys. Rev. A \textbf{64}, 063818 (2001).
\bibitem{23} M. G. A. Paris, Phys. Rev. A \textbf{62}, 033813 (2000);
             D. T. Pegg, L. S. Phillips, and S. M. Barnett, Phys. Rev. Lett. \textbf{81}, 1604 (1998);
             X. B. Zou, K. Pahlke, and W. Mathis, Phys. Rev. A \textbf{66}, 014102 (2002).
\bibitem{24} C. W. Gardiner and P. Zoller, in Quantum Noise, A Handbook of Markovian and Non-Markovian Quantum Stochastic Methods with Applications to Quantum Optics, edited by
             Hermann Haken, (2000), p. 397.                                                                                               .
\bibitem{25} X. B. Zou, K. Pahlke, W. Mathis, Phys. Rev. A \textbf{66}, 044302 (2002);
             T. Yamamoto, K. Tamaki, M. Koashi, N. Imoto, Phys. Rev. A \textbf{66}, 064301 (2002);
             X. Wang, Phys. Rev. A \textbf{64}, 012313 (2001);
             G.-P. Guo, C. F. Li, J. Li, G. C. Guo, Phys. Rev. A \textbf{65}, 042102 (2002).

\end{thebibliography}
\end{document}